\documentclass[fleqn,usenatbib,letters]{mnras}
\usepackage[T1]{fontenc}
\usepackage{newtxtext,newtxmath}
\usepackage{graphicx}
\usepackage{hyperref}
\usepackage{tabularx}
\usepackage{flafter}
\usepackage{silence}
\usepackage{pdfpages}
\graphicspath{{media/media/}{media/}}

\title[Stellar occultations by Haumea]{Predictions of Stellar Occultations by Haumea and the Event of 4 May 2026}
\author[J. L. Ortiz et al.]{
Jose L. Ortiz,$^{1}$
Nicolas Morales,$^{1}$
Antonio Oca\~na-Pastor,$^{2}$
Jos\'e M. G\'omez-Lim\'on,$^{1}$
Steve B. Howell,$^{3}$\newauthor
Francisco J. Pozuelos,$^{1}$
Pablo Santos-Sanz,$^{1}$
Yucel Kilic,$^{1}$
G\"{o}khan Y\"{u}cel,$^{4}$
Rafael Morales,$^{1}$\newauthor
and Mike Kretlow$^{5}$
\\
\\
$^{1}$Instituto de Astrof\'isica de Andaluc\'ia-CSIC, Glorieta de la astronom\'ia s/n, 18008 Granada, Spain
\\
$^{2}$I.E.S Polit\'ecnico Hermenegildo Lanz, Albond\'on 12, 18004 Granada, Spain
\\
$^{3}$NASA Ames Research Center, Moffett Field, CA 94035, USA
\\
$^{4}$Department of Space Sciences and Technologies, Faculty of Science, Akdeniz University, 07058, Antalya, T\"{u}rkiye
\\
$^{5}$Deutsches Zentrum f\"ur Astrophysik (DZA), Postplatz 1, 02826 G\"orlitz, Germany
}
\date{Accepted XXX. Received XXX; in original form XXX}
\pubyear{2026}

\begin{document}

\maketitle

\begin{abstract}
Haumea is the third-largest of the five officially recognized dwarf planets and one of the four that reside in the trans-Neptunian region. It is among the most exotic bodies in the Solar System. Because of its large distance, direct exploration by space missions is not feasible in the short term, so progress must rely on ground- and near-Earth facilities. Stellar occultations are among the most powerful tools to investigate trans-Neptunian objects. We present predictions of occultations by Haumea in the coming years for stars down to Gaia G = 21. We identify eleven relevant events through 2030. We analysed in detail the geometry of the 4 May 2026 event, including Haumea's rotation phase, known 3D shape, pole orientation, and sky-plane motion. We derive a sky-plane shadow width of 2224 $\pm$ 30 km, substantially larger than nominal assumptions and therefore highly favorable for observations. Given the star's large RUWE of 6.6, raising concerns about the detectability of the event, we also performed a reliability analysis. Speckle imaging reveals a companion at $\sim$0.12 arcsec and $\Delta m \approx 3.1$; this companion is also expected to be occulted, and shifts the nominal main-star path prediction on Earth by about 8 mas.

\end{abstract}


\section{Introduction}\label{introduction}

Haumea is one of the most remarkable objects in our Solar System. It is one of the five officially recognized dwarf planets by the International Astronomical Union, four of which lie in the trans‑Neptunian region, and Haumea is among them. It is an extremely fast‑spinning body for its size, with a rotational period below 4 hours, and it exhibits a highly elongated triaxial shape \citep{Rabinowitz2006ApJ_HaumeaLightcurve}. It also possesses two known satellites and a group of bodies with similar orbital characteristics \citep{Brown2007NatureFamily}, whose origin is still debated. Beyond all these features, Haumea is also an exotic object because it hosts a ring. Besides Haumea, only three other non-planetary Solar System bodies are currently known to have rings: the Centaur Chariklo \citep{BragaRibas2014NatureChariklo}, the Centaur Chiron \citep{Ortiz2023Chiron,Pereira2025Chiron}, and the trans‑Neptunian object Quaoar \citep{Morgado2023NatureQuaoarRing,Pereira2023AandA_TwoRingsQuaoar}. The discovery of Haumea's ring and other key findings resulted from a stellar occultation \citep{Ortiz2017NatureHaumeaRing}.

But the 2017 occultation by Haumea also revealed that its size was significantly larger than previously thought. As a consequence, Haumea's inferred density decreased from the then‑accepted value of around 2700 kg m$^{-3}$, to approximately 2000 kg m$^{-3}$, consistent with the densities of other large bodies such as Pluto and Triton \citep{Ortiz2017NatureHaumeaRing}. However, because this occultation provided only a single geometric cross‑section, assumptions had to be made regarding the ratio of Haumea's two largest axes, using constraints derived from the amplitude of its rotational light curve. The precise relationship between light‑curve amplitude and axis ratio depends on the photometric model adopted to describe the surface scattering law. Although the pole orientation of the ring, assumed to be equatorial, could in principle allow reconstruction of the full 3‑D shape without additional assumptions, this requires accurate knowledge of Haumea's rotational phase at the time of the occultation \citep[e.g.,][]{Dunham2019ApJ_HaumeaInternal}. Furthermore, the ring plane may precess slightly due to Haumea's significant J2 component of its gravitational potential (because of Haumea's elongated shape) or may exhibit small eccentricities, both of which complicate the determination of the ring pole. Recent dynamical modelling suggests that long-term precession of Haumea's ring plane should be very small: when the effects of the body's triaxial figure, satellite perturbations, and collisional damping are taken into account, the ring is expected to remain close to the equatorial plane, although non-negligible particle eccentricities may still be sustained \citep{Marzari2020HaumeaRingDynamics}. As a result, Haumea's triaxial shape can still be refined, and the same is true for its density, through additional occultation observations.

An accurate density estimate is essential for modeling Haumea's interior and determining whether it is in hydrostatic equilibrium, an issue that remains unresolved although a differentiated body may be a solution \citep{Dunham2019ApJ_HaumeaInternal}. Besides, if a sufficient number of occultation chords are obtained in future events, Haumea's topography could be reconstructed in detail, similarly to what has recently been achieved for the large TNO (307261) M\'ani, provisionally designated 2002 MS4 \citep{Rommel2023}. Moreover, much remains to be learned about the ring. For instance, if the ring particles are small, on the order of a few microns, the opacity can depend strongly on wavelength \citep{Kalup2025RingOpacity}, allowing the particle‑size distribution to be constrained through multi‑wavelength occultation observations \citep[e.g.,][]{SantosSanz2025Char}. Additional faint satellites or narrower rings may be detected, and tighter upper limits on any possible atmosphere could be derived. The large satellite Hi`iaka may also induce precession effects on the ring, which merit investigation. \citet{Marzari2020HaumeaRingDynamics} showed that Hi`iaka likely does not have a strong effect on the nodal precession of the rings since it is coplanar with the rings to within $\sim$1$^{\circ}$. Namaka, although smaller, can induce a larger nodal precession effect because of its higher $\sim$13$^{\circ}$ inclination. On the other hand, apsidal precession may be dominated by Hi`iaka.

For all these reasons, the scientific motivation to observe stellar occultations by Haumea is very high. Unfortunately, such occultations are rare. In this work, we conducted an extensive search for occultations of stars as faint as magnitude 21 through the year 2031, in order to identify all possible opportunities and highlight those with the greatest potential. These predictions are presented in the first section. The two events with the highest expected scientific return are discussed, with particular emphasis on the most promising event, that of 4 May 2026, in section 3. Finally, conclusions and a summary are provided.

\section{Predictions}\label{predictions}

Predictions of stellar occultations by Haumea involving stars down to magnitude 21 in the Gaia G band were carried out for the interval 2024--2031 (inclusive). To do this, we extracted the positions of all Gaia DR3 stars located within the sky region traversed by Haumea through 2031. We then searched for close angular approaches between Haumea and those stars as seen from Earth, retaining only those cases with the potential to produce a ground‑track (shadow path) on any region of the Earth. The source of Haumea's ephemerides was the JPL Horizons system. Several events with some observability potential were identified (see table 1). Maps showing the predicted shadow paths and auxiliary information for these events are provided as Supplementary Material (occultation maps for events 1--11). 

Note that a sky-plane shadow width of 1380 km was adopted in those computations. Because the instantaneous projected figure of Haumea on the sky is elliptical, the apparent sky-plane shadow width depends on Haumea's sky‑plane motion vector relative to the pole position angle and on its rotational phase. An average effective value would be around 1700 km, but at rotational minimum the projected diameter reaches 1380 km. We therefore preferred to use this very conservative value for the nominal predictions. The best event in terms of stellar brightness and broad observability across large terrestrial areas is that of 2026 May 4. Because the shadow track spans several thousand kilometers across three continents and the involved star has Gaia G = 14.7, the event can potentially yield excellent observations, as adverse weather is unlikely to affect all accessible locations simultaneously. For context, the star occulted in 2017 by Haumea had G = 17.8, more than 3 magnitudes fainter, implying that the 2026 May 4 occultation can be observed with significantly smaller instruments than those used by \citet{Ortiz2017NatureHaumeaRing}. However, see the considerations in the next paragraph regarding the prediction of the shadow path and other aspects.

The event on 2027 July 29 is also encouraging, because the occultation star has V = 15.7 and the ground track crosses a sufficiently large and populated part of the world, although the event occurs in twilight for most locations. One additional event in 2030 involves a star of V = 16.4, but the corresponding ground track is less favorable. All of these events are, in principle, observable with small telescopes of 0.2 m aperture, as suggested by simulations carried out under average atmospheric conditions and using typical observing parameters.

The remaining events involve fainter stars in the V = 18.3--20.4 range. Given that Haumea itself is around V = 17.2, the combined, unresolved source is only slightly brighter than Haumea alone; thus, the occultations will produce brightness drops of only \textasciitilde{}0.4 mag to \textasciitilde{}0.04 mag on a \textasciitilde17 mag source. Most of these fainter‑star events will require medium to large telescopes to detect the flux decrements with sufficiently short cadence to determine accurate ingress and egress times, aiming for \textasciitilde{}10 km accuracy in the reconstructed chord lengths. However, since all the involved stars are brighter at longer wavelengths, unfiltered or R‑band observations may deliver larger effective brightness drops, which in turn could make detections feasible with relatively modest equipment in the 0.4 m class. The long duration of the events at around 90s on average also helps for the detectability of the events.

\section{The 2026 May 4 Event}\label{the-2026-may-4th-event}

For the 2026 May 4 event, which is the one with the highest anticipated scientific return, there are some aspects that deserved closer scrutiny. For instance, the Renormalized Unit Weight Error (RUWE) value of the occultation star in the Gaia DR3 catalog is exceptionally high (RUWE = 6.6). Such a value clearly indicates problems with the star's astrometric solution. Astrometric errors of several tens of milliarcseconds can significantly impact the predicted shadow path on Earth and therefore affect the observability of the event. High RUWE values (typically RUWE \textgreater{} 1.4) are often associated with unresolved binarity or multiplicity (e.g., \citealt{CastroGinard2024AandA_RUWEBinaries}) although other aspects can give rise to high RUWE. Thus, this deserved closer investigation.

Based on the Gaia colors and additional data retrieved through the VizieR service, the star appears to be of G spectral type. The $B_{\mathrm{p}}-R_{\mathrm{p}}=0.83497$ color and the $T_{\mathrm{eff}}=5577$ K from Gaia, together with spectra obtained using the CAFOS instrument at the Calar Alto Observatory, are fully consistent with a G‑type classification. Given its brightness, the star's approximate distance should be around 700 pc. Any companion stars would have to be either also G‑type or sufficiently faint so as not to contribute significantly to the combined optical spectrum.

To investigate the origin of the high RUWE value, we arranged speckle imaging observations to search for close companions. These observations were carried out on 13 April 2025 with the 8.1‑m Gemini North Telescope using the \textquotesingle Alopeke dual‑channel speckle imager \citep{Scott2021FrASS_AlopekeZorro,Howell2011AJ_KeplerSpeckle,Horch2006RMxAC_SpeckleStatus,Howell2024AJ_MFBDspeckle}. Two narrowband filters were used: one centered at 562 nm (FWHM = 54 nm) and another at 832 nm (FWHM = 40 nm). \textquotesingle Alopeke operates by acquiring thousands of very short exposures (10--60 ms), each effectively freezing atmospheric distortion from turbulence (as this is the rate at which an isoplanatic patch of the atmosphere varies with time; \citealt{Fried1966JOSA_r0}), and subsequently combining these exposures through Fourier reconstruction techniques to obtain diffraction‑limited images.

During the observing run, several calibration binary stars with well‑established orbital solutions, separations, position angles, and magnitude differences were observed to determine the plate scale and camera orientations. The measured pixel scales were 0.009826 arcsec/pixel for the blue channel and 0.010451 arcsec/pixel for the red channel. The corresponding detector orientations were --89.37° (blue) and --89.11° (red). Further details on the technique and the instrument calibration can be found in Howell et al. (2024).

\begin{figure}
  \centering
  \includegraphics[width=0.92\columnwidth]{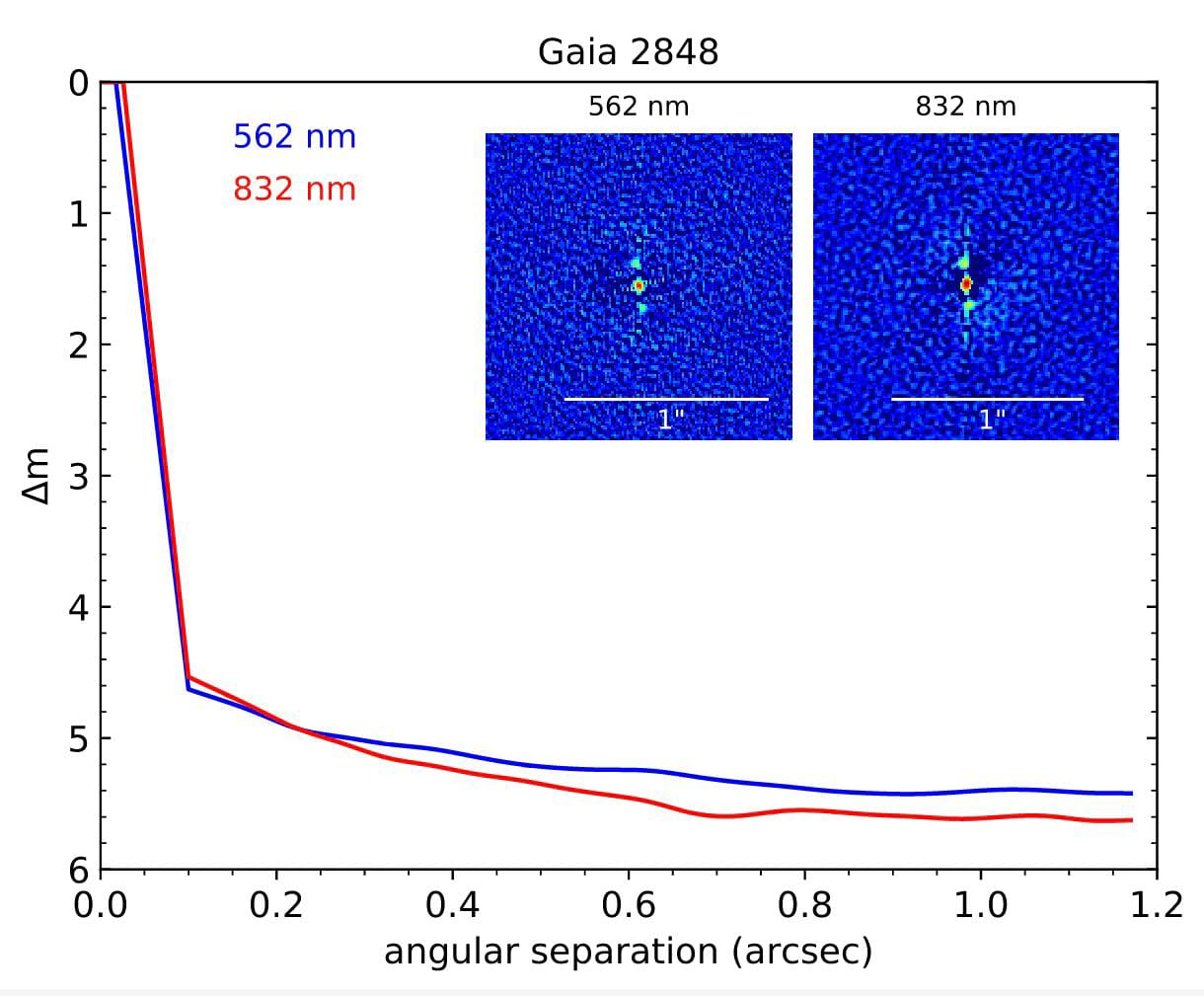}
  \caption{Plot of the contrast in magnitudes for the occultation star versus angular separation. The inserts show the speckle combined images at the two observing wavelengths. A companion of the main star at 0.12 arcsec with a brightness of 3.1 mag fainter than the primary is clearly seen. The background contrast limits achieved tend to be shallow close to the star, only reaching their full contrast depth near separations of \textasciitilde{}0.1 and beyond (see Howell \& Furlan 2022). Speckle interferometric Fourier reconstructions contain a 180° ambiguity (a ghost) for detected close companions, which is the case here.}
  \label{fig:contrast}
\end{figure}

The observations revealed the presence of a companion at a separation of 0.10 ± 0.02 arcsec at 832 nm and 0.14 ± 0.02 arcsec at 562 nm, with a magnitude difference of $\Delta m$ = 3.1 ± 0.3 (see Fig.~\ref{fig:contrast}). The position angle was measured to be 7 ± 2 degrees, or its supplementary value (due to the intrinsic 180° ambiguity inherent to speckle interferometric reconstructions). Fig.~\ref{fig:contrast} presents the resulting contrast curves: the background contrast reaches meaningful depth only beyond separations of \textasciitilde{}0.1 arcsec \citep{HowellFurlan2022FrASS_GeminiSpeckle}. The typical ghost solution (mirror ambiguity) is also evident in the data.

This detected companion affects the occultation prediction because the system's photocenter is shifted by approximately 8 milliarcseconds toward the companion relative to the primary star's nominal position in Gaia DR3. This is not a severe shift given the much larger angular size of the shadow. Bispectral analysis of the phase information contained in the images \citep{Horch2011AJ_DSSI} can be used to select the correct quadrant; in our case it favoured the northern location for the companion star, although the southward solution cannot be completely discarded.

Another consequence of our finding of a companion at \textasciitilde0.12 arcsec is that this star will also be occulted, with a shadow path most likely favorable for the south of Africa, although the north of Europe solution remains possible. As the companion is \textasciitilde3.1 mag fainter than the primary, the occultation will be similar to that in 2017 \citep{Ortiz2017NatureHaumeaRing} and will cause brightness drops detectable with more powerful telescopic resources than for the primary. But this enhances the detectability of the event and the potential scientific return.

\begin{figure}
  \centering
  \includegraphics[width=0.92\columnwidth]{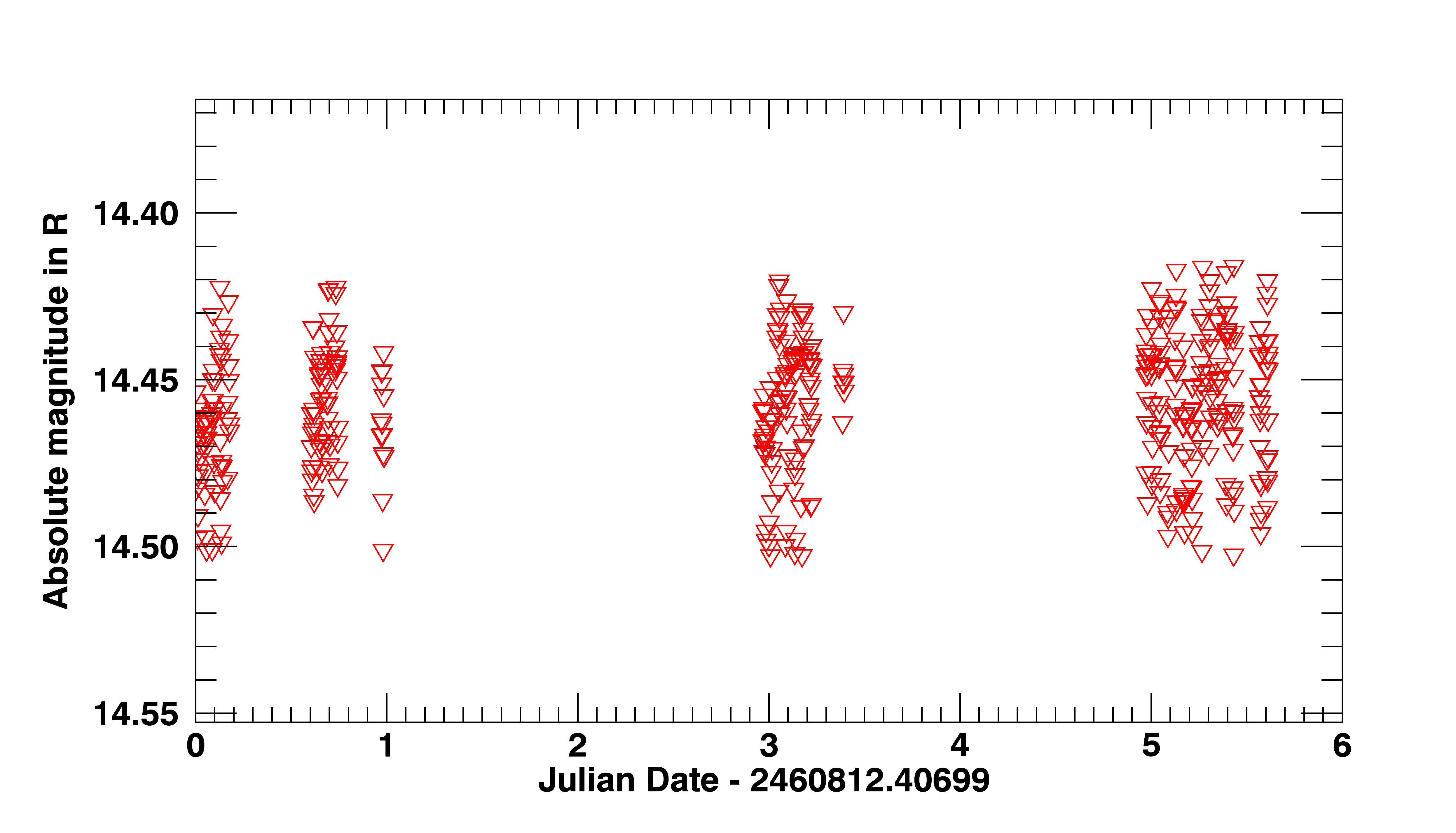}\\[0.8em]
  \includegraphics[width=0.92\columnwidth]{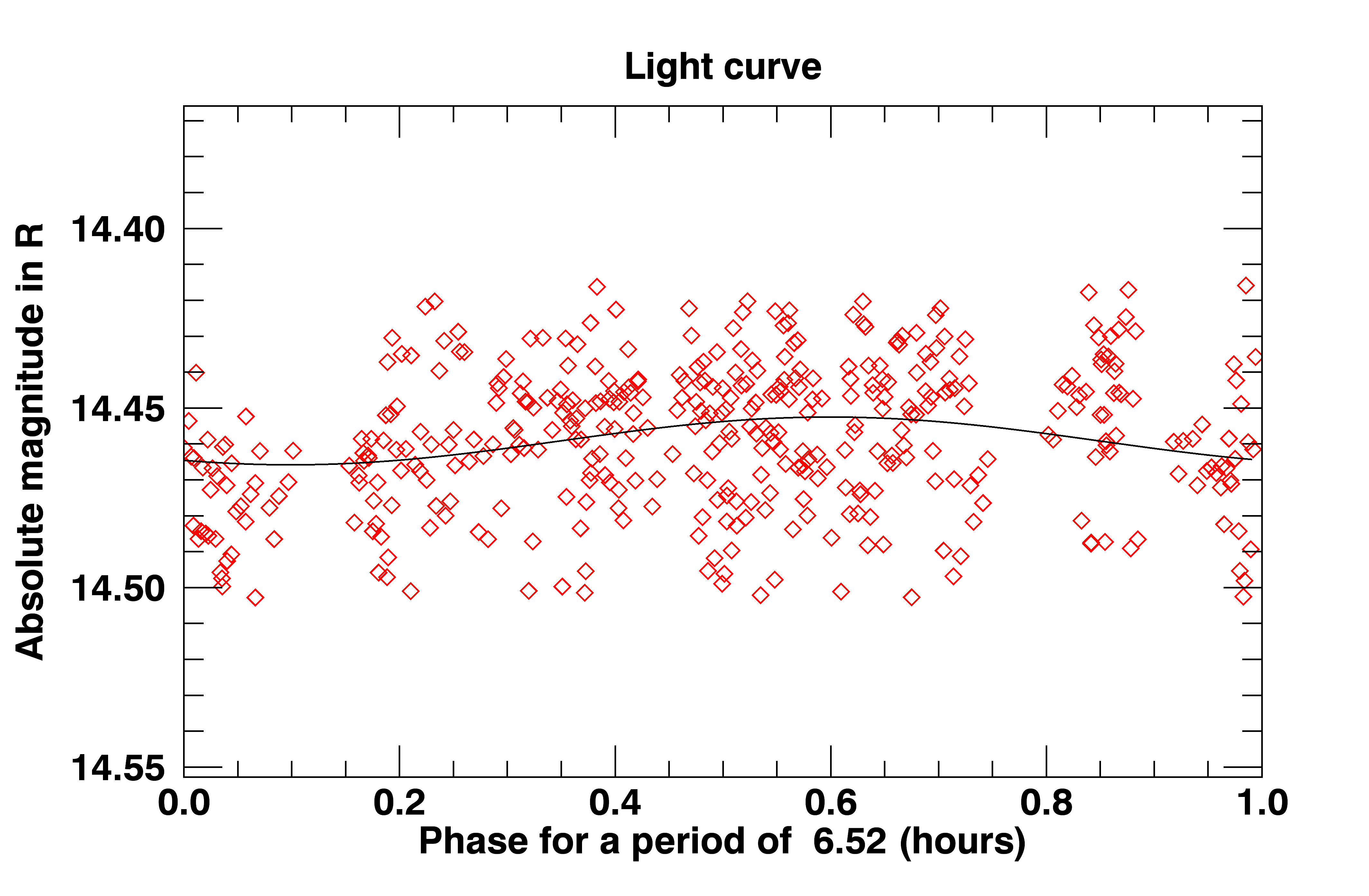}
  \caption{LCO photometric monitoring of star UCAC4 524-056397. Top panel: photometry versus time obtained with LCO 0.35m telescopes. Bottom panel: light curve folded to a period of 6.52 hours. The diamonds represent the observed brightness as a function of phase, using the first observation as the reference for zero phase. The continuous line is the fitted curve, with a peak-to-valley oscillation of 0.013 mag.}
  \label{fig:photometry}
\end{figure}

If the detected speckle companion is gravitationally bound to the primary star (rather than being a chance line‑of‑sight background source), its separation would correspond to a physical distance of more than 1000 astronomical units, implying an orbital period too long to have produced Gaia's high RUWE value through photocenter motion. Using the RUWE simulator developed by \citet{CastroGinard2024AandA_RUWEBinaries}, we find that such a wide binary configuration cannot reproduce the observed RUWE for a G‑type primary. A more probable explanation is that Gaia resolved the pair in some scan angles of Gaia but not in others (their separation lies close to Gaia's diffraction-limited resolution of \textasciitilde0.15 arcsec) or that there is flux variability in one or both components, which can perturb the RUWE normalization \citep{CastroGinard2024AandA_RUWEBinaries} or shift the photocenter. For instance, variations as small as 0.1 mag in the secondary (\textasciitilde{}0.01 mag in the combined unresolved flux) could induce photocenter motions on the order of \textasciitilde1 mas that would be absorbed as noise in Gaia's astrometric solution. The astrometric noise in Gaia DR3 for this star is listed as 1.3 mas, which would be consistent with the picture of variability at the 1 percent level in the combined source. The source is not listed as a variable in Gaia DR3, but the low amplitude and other aspects can result in variability being undetected with the Gaia DR3 pipeline \citep{Eyer2023AandA_GaiaDR3Variability}. Alternatively, an additional unresolved much closer companion could exist. Exploring parameter space with the Castro‑Ginard (2024) formalism, we find that such hypothetical companions can explain a RUWE of 6.6 but would generate photocenter shifts well below 1 mas, and therefore would have no appreciable effect on the predicted shadow path.

\begin{table*}
\centering
\scriptsize
\setlength{\tabcolsep}{3.5pt}
\renewcommand{\arraystretch}{1.08}
\caption{Stellar occultation opportunities (2026-2030)}
\label{tab:occultation_mv}
\begin{tabularx}{\textwidth}{l l >{\raggedright\arraybackslash}X c c c c c}
\hline
\textbf{Occultation Date} & \textbf{Approx. Time (UT)} & \textbf{Star Coordinates (RA, Dec)} & \textbf{Mv} & \textbf{RUWE} & \textbf{Max Dur. (s)} & \textbf{Lun. Dist. ($^\circ$)} & \textbf{Illum. (\%)} \\
\hline
2026 May 04 & 20:12 to 20:22 & RA 14 40 58.5, Dec +14 40 26 & 14.7 & 6.6 & 66 & 55 & 91 \\
2027 Jul 29 & 02:10 to 02:26 & RA 14 40 58.5, Dec +14 03 13 & 19.4 & 1.0 & 93 & 134 & 25 \\
2027 Jul 29 & 13:52 to 14:08 & RA 14 42 58.4, Dec +13 55 61 & 15.7 & 1.0 & 92 & 130 & 20 \\
2028 Jan 02 & 15:32 to 15:44 & RA 14 50 48.3, Dec +12 58 38 & 18.3 & 1.0 & 32 & 5.1 & 68 \\
2028 May 16 & 12:09 to 12:18 & RA 14 47 31.3, Dec +14 02 07 & 18.4 & 1.0 & 32 & 5.0 & 44 \\
2029 Jan 02 & 01:08 to 01:19 & RA 14 54 24.3, Dec +12 38 09 & 19.8 & 1.0 & 65 & 98 & 98 \\
2029 Feb 15 & 22:30 to 22:46 & RA 14 58 35.1, Dec +12 58 25 & 20.2 & 1.0 & 94 & 128 & 6 \\
2030 Jan 13 & 10:20 to 10:35 & RA 14 58 32.3, Dec +12 46 36 & 20.4 & 1.1 & 81 & 148 & 67 \\
2030 Mar 04 & 14:08 to 14:23 & RA 14 59 01.8, Dec +12 47 09 & 20.2 & 1.1 & 82 & 121 & 0 \\
2030 May 12 & 19:02 to 19:12 & RA 14 55 13.1, Dec +13 19 42 & 20.1 & 0.9 & 63 & 58 & 72 \\
2030 May 24 & 14:15 to 14:27 & RA 14 54 25.4, Dec +13 21 11 & 16.4 & 0.9 & 65 & 114 & 46 \\
\hline
\end{tabularx}
\vspace{2pt}
\begin{minipage}{\textwidth}
\footnotesize
\textit{Notes.} ``Approx. Time (UT)'' indicates the time interval during which Haumea's shadow crosses the Earth. The listed coordinates correspond to the occultation epoch (i.e., stellar positions propagated to the event date). ``Max Dur. (s)'' is the maximum theoretical duration at central chord, ``Lun. Dist.'' is the angular distance to the Moon, and ``Illum.'' is the lunar illuminated fraction.
\end{minipage}
\end{table*}

To investigate the photometric variability we made observations with the Las Cumbres Observatory (LCO) 0.35m Delta Rho 350 telescopes equipped with QHY600 sCMOS cameras (see instrument details \url{https://lco.global/observatory/instruments/qhy600-delta-rho-350/}). We used the r\_sloan filter for 5 nights to monitor the star brightness to look for potential short-term variability. The observations were carried out from 2025/05/17 to 2025/05/22 using 40s exposures. The Tenerife, Haleakala, MacDonald and Siding Spring sites could make observations. There was no continuous monitoring from the first night to the last night without gaps, but there was almost complete 24-h coverage in the first and last nights. Synthetic aperture photometry was carried out and instrumental magnitudes converted to the Johnson R system were obtained. Fig.~\ref{fig:photometry} illustrates the time coverage and the derived magnitude. The errors of the individual measurements are only 0.01 mag on average. A Lomb-Scargle periodogram analysis shows a prominent peak at 6.52h with a confidence level well above the 99\% level and the light curve folded to that period has a peak to valley amplitude of 0.013 mag (bottom panel of Fig.~\ref{fig:photometry}). This seems consistent with the possibility that the high RUWE of the star could come from variability in the secondary or in the primary.

We also investigated whether the proper motion values in Gaia DR3 might be affected by the high RUWE and result in a poor propagation of the position of the star for the occultation time. However, the Gaia DR3 values are consistent with those listed in the "Hot Stuff for One Year" (HSOY) catalog \citep{Altmann2017HSOY}, which is based mostly on the PPMXL catalog \citep{Roeser2010PPMXL}, combined with Gaia DR2 positions. HSOY lists proper motions of -7.724 mas/yr (RA) and -8.740 mas/yr (Dec), for the star, while GaiaDR3 gives -7.394 ± 0.170 mas/yr and -8.833 ± 0.139 mas/yr. Propagating the positions in the catalogs to the epoch of the occultation yields a difference of only 8 mas in declination (+14 40 25.612 for HSOY vs. +14 40 25.604 for Gaia DR3), while the right ascension values agree to the milliarcsecond level (14:40:58.4947 in both cases). Thus, even if Gaia's proper motions were slightly biased, the impact on the prediction would be minimal.

\begin{figure*}
  \centering
  \includegraphics[width=\textwidth,height=0.82\textheight,keepaspectratio]{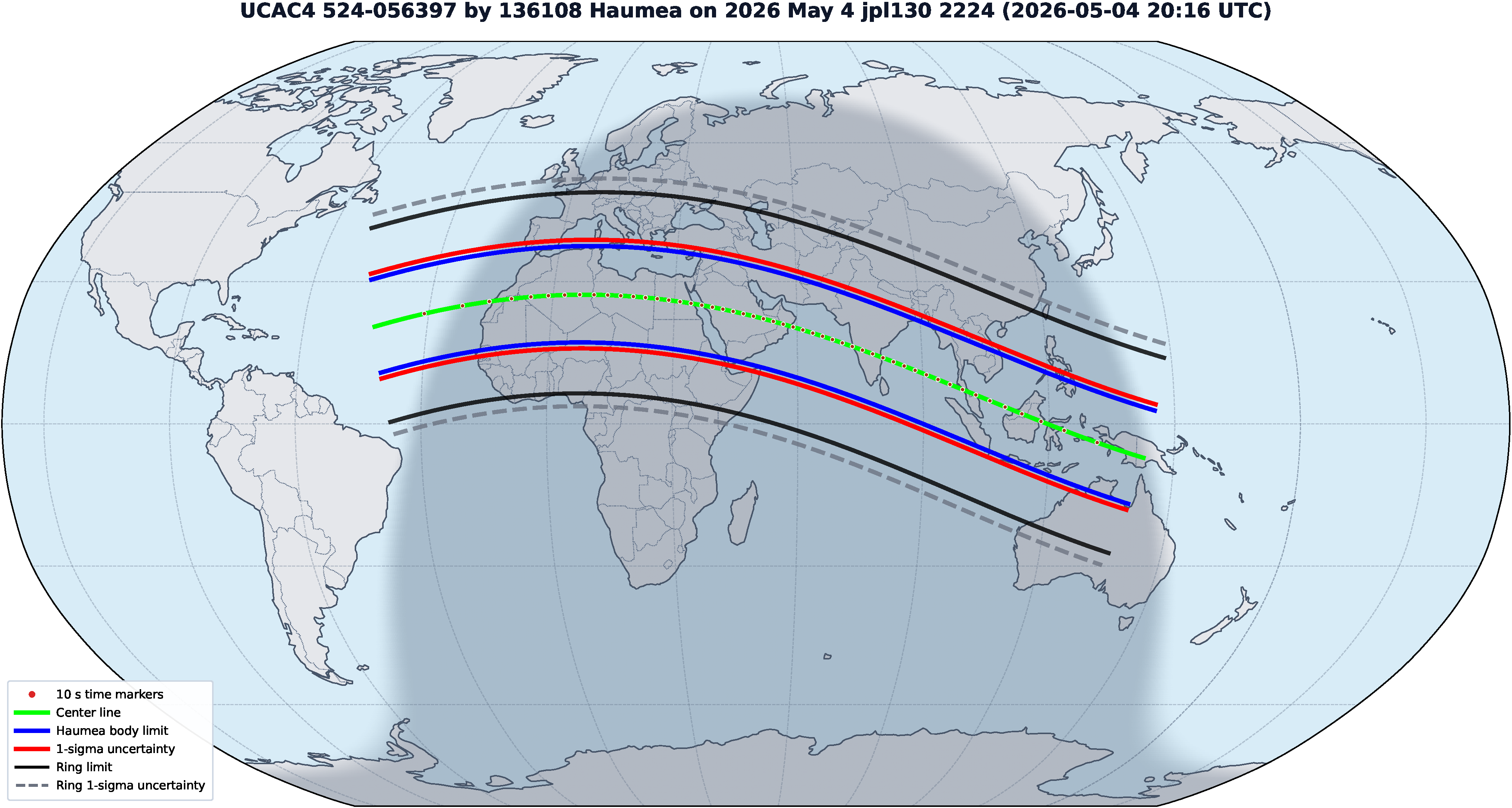}
  \caption{Predicted geometry of the 2026 May 4 event showing the broader ring-occultation shadow region relative to the main-body shadow. A zoomable version of the map is available at \url{https://opop.obspm.fr/media/data/chords/156331/Haumea_4th_May_2026_occultation_map_IAA-CSIC.html}. Observations of this event can be reported through the occultation portal at \url{https://opop.obspm.fr/create_report/2518/}.}
  \label{fig:ring-shadow-2026}
\end{figure*}

Another aspect to consider for this occultation is the rotation phase of Haumea. Because Haumea is highly elongated and rotates rapidly, its projected area changes substantially over a rotation cycle, and consequently the sky-plane shadow width also varies. Therefore, it is important to know Haumea's rotational phase at the time of the occultation. We computed this using a light curve obtained in 2022 with the 1.5‑m telescope at Sierra Nevada Observatory, and then propagated the phase using the rotation period of 3.915341 ± 0.000005 hours \citep{Ortiz2017NatureHaumeaRing}. At the time of the occultation, the rotational phase will be only 25 degrees from maximum brightness (corresponding to maximum projected size). Using the spin‑axis orientation derived in \cite{Ortiz2017NatureHaumeaRing} and the 3D shape there, we estimate a shadow width of 2224 ± 30 km. This is a favorable result, as it indicates a much wider shadow path than the conservative nominal value of 1380 km used for initial predictions and highly enhances the detectability of the main‑body occultation (see Fig.~\ref{fig:ring-shadow-2026}). A recent study \citep{Arxiv2026HaumeaDeformation} also suggests that Haumea may possess an equatorial deformation large enough to be detectable in forthcoming occultations, which further strengthens the scientific interest of well-sampled events such as that of 2026 May 4.

\begin{figure*}
  \centering
  \includegraphics[width=\textwidth,height=0.82\textheight,keepaspectratio]{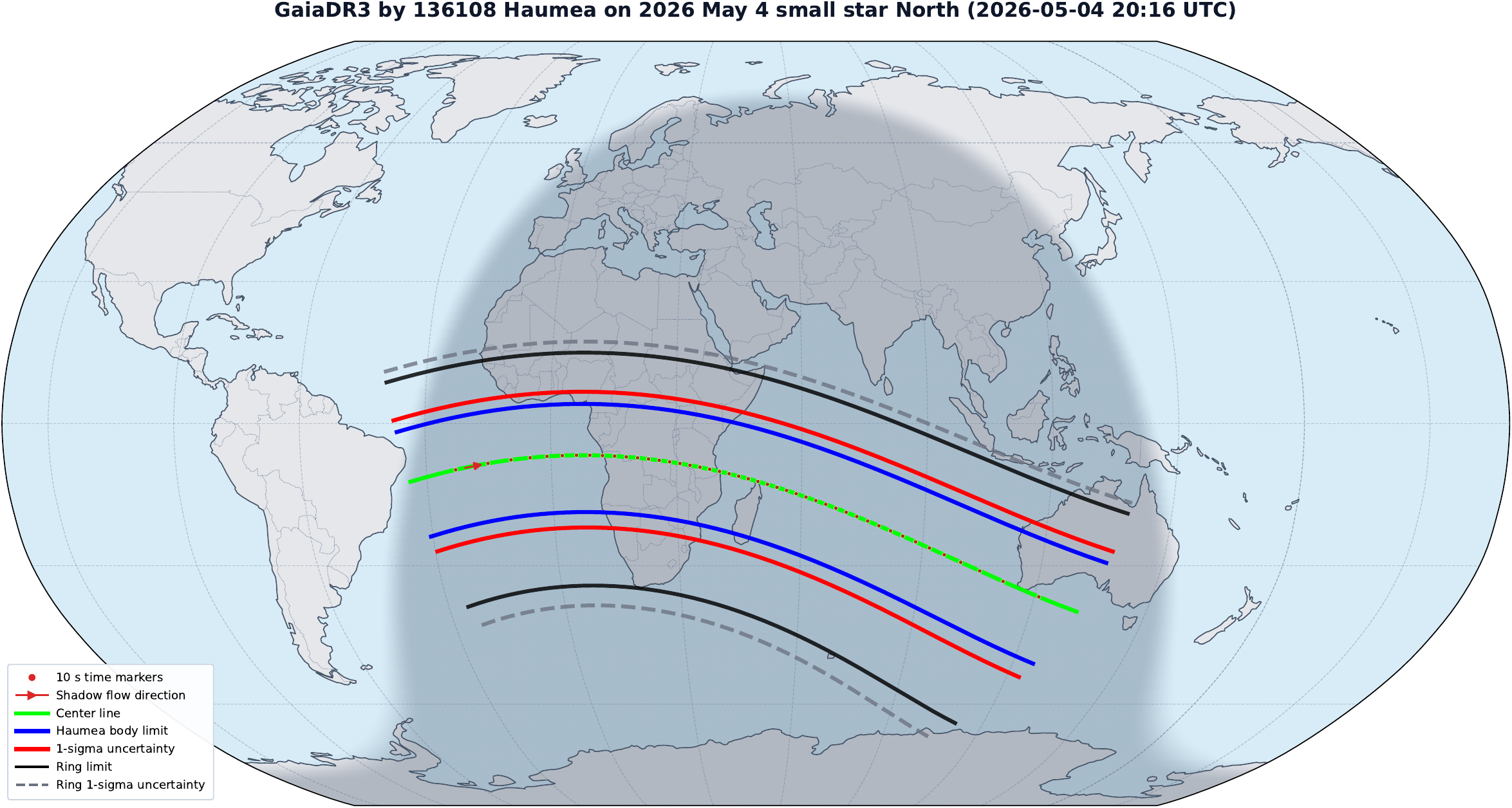}
  \caption{Same as Fig.~\ref{fig:ring-shadow-2026} but for the companion star in its preferred position. The arrow indicates the motion of the shadow.}
  \label{fig:ring-shadow-2026-companion}
\end{figure*}

Another factor to consider is that Haumea possesses a ring system, meaning that the star's occultation by the ring may be observable across a considerably wider region of Earth than the main‑body occultation itself (see Fig.~\ref{fig:ring-shadow-2026}). The shadow path corresponding to the companion star in its preferred position for both the main body and the ring is shown in Fig.~\ref{fig:ring-shadow-2026-companion}.

Regarding the effect of Haumea's motion around the barycenter of the Haumea--Hi`iaka system, we should point out that our predictions use the most recent JPL Horizons orbit and this corresponds to the barycenter of the system: Since most astrometric measurements assimilated into Horizons correspond to unresolved images of the Haumea system, the positions refer to the system\textquotesingle s photocenter, which orbits the barycenter. Over long data arcs this motion averages out, so the fitted heliocentric orbit effectively represents the system barycenter. However, given current mass estimates \citep{Proudfoot2024PSJ_HaumeaJ2}, the barycenter would lie roughly 6 mas away from Haumea toward Hi`iaka. The resulting offset depends on Hi`iaka's orbital phase. For the May 4 event,according to the Miriade service\footnote{\url{https://vo.imcce.fr/webservices/miriade}}, Hi`iaka is predicted to be at ($\Delta$RA, $\Delta$Dec) = (-0.03", +1.02"), which corresponds to corrections of only +0.1 mas in right ascension and -4.7 mas in declination when converting barycentric coordinates to Haumea's position. These corrections produce small changes (about 166 km cross-track) in the predicted ground track compared to the width of the path. Note that both Hi`iaka and Namaka are angularly so far from Haumea at the date of the occultation that the corresponding shadow paths lie outside the Earth.

Observations of the event can be reported via the occultation portal \citep{Kilic2020OCCPORTAL} at the following entry: \url{https://opop.obspm.fr/create_report/2518/} where zoomable maps are also available.

\section{Summary}\label{summary}

Stellar occultations by Haumea are not frequent, but by extending our predictions down to very faint stars (as faint as magnitude 21), we have identified a set of eleven good opportunities that are particularly valuable for studying this exotic dwarf planet and its ring system until the end of 2030. Apart from refining the 3D shape of Haumea, its density, the topography level and some ring features, these events carry the potential for discovering additional rings or small satellites around Haumea and for placing stronger constraints on any possible atmosphere. Among the selected occultations, the event with the highest observability and expected scientific return involves a relatively bright star that nevertheless has a large RUWE value in the Gaia catalog, an indication of potential issues with the star's astrometry and a potential source of uncertainty in the predicted ground track. We presented follow‑up observations of this star. The speckle results clearly reveal a companion at 0.10--0.14 arcsec, below Gaia's typical resolving limit, which shifts the predicted path by approximately 8 mas. This is roughly four times smaller than the angular size of Haumea's main body, so the overall observability prospects remain strong.

Although the 180° ambiguity in the speckle position angle is not fully resolved, the northward companion star is currently preferred. In either case, the occultation can still be recorded from numerous sites, and an event by the fainter companion is expected most likely over southern Africa, with a secondary possible solution over northern Europe. Given the large angular extent of the ring system, the ring occultation will be observable over a wide fraction of the Earth. After the occultation is observed we will also be able to learn more about the situations that result in large RUWE in the Gaia DR3 catalog and we will be able to make progress on the knowledge of stellar binarity through the combination of Gaia data and stellar occultation results. We also presented other occultation opportunities, including visibility maps, to assist with detailed observation planning.

\section*{Supplementary material}

Supplementary material is available online as a single PDF file (\texttt{supplementary.pdf}) containing the occultation prediction maps and geometry plots for events 1--11.

\section*{Acknowledgements}
This work was supported by multiple funding agencies and institutions.
It was partly funded by the Spanish projects PID2020-112789GB-I00 (AEI)
and Proyecto de Excelencia de la Junta de Andalucía PY20-01309.
J.L.O., P.S.-S., and N.M. acknowledge financial support from the
Severo Ochoa grant CEX2021-001131-S
(MCIN/AEI/10.13039/\allowbreak 501100011033).
P.S.-S. also acknowledges support from the Spanish I+D+i project
PID2022-139555NB-I00 (TNO-JWST), funded by
MCIN/AEI/10.13039/\allowbreak 501100011033.
We thank A. Alvarez-Candal for comments and suggestions on the manuscript.
%%%%%%%%%%%%%%%%%%%%%%%%%%%%%%%%%%%%%%%%%%%%%%%%%%

\section*{Data Availability}
The data underlying this article will be shared on reasonable request to the corresponding author.

%%%%%%%%%%%%%%%%%%%% REFERENCES %%%%%%%%%%%%%%%%%%
\bibliographystyle{mnras}
\bibliography{ref}

@article{BragaRibas2014NatureChariklo,
  author  = {Braga-Ribas, F. and Sicardy, B. and others},
  title   = {A ring system detected around the Centaur (10199) Chariklo},
  journal = {Nature},
  year    = {2014},
  volume  = {508},
  pages   = {72--75},
  doi     = {10.1038/nature13155}
}

@article{Brown2007NatureFamily,
  author  = {Brown, M. E. and Barkume, K. M. and Ragozzine, D. and Schaller, E. L.},
  title   = {A collisional family of icy objects in the Kuiper belt},
  journal = {Nature},
  year    = {2007},
  volume  = {446},
  number  = {7133},
  pages   = {294--296},
  doi     = {10.1038/nature05619}
}

@article{CastroGinard2024AandA_RUWEBinaries,
  author  = {Castro-Ginard, A. and Penoyre, Z. and Casey, A. R. and Brown, A. G. A. and Belokurov, V. and Cantat-Gaudin, T. and others},
  title   = {Gaia DR3 detectability of unresolved binary systems},
  journal = {Astronomy \& Astrophysics},
  year    = {2024},
  volume  = {688},
  pages   = {A1},
  doi     = {10.1051/0004-6361/202450172}
}

@article{Dunham2019ApJ_HaumeaInternal,
  author  = {Dunham, E. T. and Desch, S. J. and Probst, L.},
  title   = {Haumea’s Shape, Composition, and Internal Structure},
  journal = {The Astrophysical Journal},
  year    = {2019},
  volume  = {877},
  number  = {2},
  pages   = {131},
  doi     = {10.3847/1538-4357/ab13b3}
}

@article{Eyer2023AandA_GaiaDR3Variability,
  author  = {Eyer, L. and Audard, M. and Holl, B. and others},
  title   = {Gaia Data Release 3: Summary of the variability processing and analysis},
  journal = {Astronomy \& Astrophysics},
  year    = {2023},
  volume  = {674},
  pages   = {A13},
  note    = {DR3 variability summary}
}

@article{Fried1966JOSA_r0,
  author  = {Fried, D. L.},
  title   = {Optical Resolution Through a Randomly Inhomogeneous Medium for Very Long and Very Short Exposures},
  journal = {Journal of the Optical Society of America},
  year    = {1966},
  volume  = {56},
  number  = {10},
  pages   = {1372--1379},
  doi     = {10.1364/JOSA.56.001372}
}

@inproceedings{Horch2006RMxAC_SpeckleStatus,
  author    = {Horch, E. P.},
  title     = {The Status of Speckle Imaging in Binary Star Research},
  booktitle = {Revista Mexicana de Astronom{\'\i}a y Astrof{\'\i}sica, Serie de Conferencias},
  year      = {2006},
  volume    = {25},
  pages     = {79--82}
}

@article{Horch2011AJ_DSSI,
  author  = {Horch, E. P. and van Altena, W. F. and Howell, S. B. and Sherry, W. H. and Ciardi, D. R.},
  title   = {Observations of Binary Stars with the Differential Speckle Survey Instrument. III. Measures below the Diffraction Limit of the WIYN Telescope},
  journal = {The Astronomical Journal},
  year    = {2011},
  volume  = {141},
  number  = {6},
  pages   = {180},
  doi     = {10.1088/0004-6256/141/6/180}
}

@article{Howell2011AJ_KeplerSpeckle,
  author  = {Howell, S. B. and Everett, M. E. and Sherry, W. and Horch, E. and Ciardi, D. R.},
  title   = {Speckle Camera Observations for the NASA Kepler Mission Follow-up Program},
  journal = {The Astronomical Journal},
  year    = {2011},
  volume  = {142},
  pages   = {19},
  doi     = {10.1088/0004-6256/142/1/19}
}

@article{HowellFurlan2022FrASS_GeminiSpeckle,
  author  = {Howell, S. B. and Furlan, E.},
  title   = {Speckle Interferometric Observations With the Gemini 8-m Telescopes: Signal-to-Noise Calculations and Observational Results},
  journal = {Frontiers in Astronomy and Space Sciences},
  year    = {2022},
  volume  = {9},
  pages   = {871163},
  doi     = {10.3389/fspas.2022.871163}
}

@article{Howell2024AJ_MFBDspeckle,
  author  = {Howell, S. B. and Martinez, A. O. and Hope, D. A. and Ciardi, D. R. and Jefferies, S. M. and Baron, F. R. and Lund, M. B.},
  title   = {High-contrast, High-angular-resolution Optical Speckle Imaging: Uncovering Hidden Stellar Companions},
  journal = {The Astronomical Journal},
  year    = {2024},
  volume  = {167},
  pages   = {258},
  doi     = {10.3847/1538-3881/ad3df2}
}

@article{Morgado2023NatureQuaoarRing,
  author  = {Morgado, B. E. and Sicardy, B. and Braga-Ribas, F. and Ortiz, J. L. and Salo, H. and Vachier, F. and Desmars, J. and Pereira, C. L. and Santos-Sanz, P. and Sfair, R. and de Santana, T. and others},
  title   = {A dense ring of the trans-Neptunian object Quaoar outside its Roche limit},
  journal = {Nature},
  year    = {2023},
  volume  = {614},
  pages   = {239--243},
  doi     = {10.1038/s41586-022-05629-6}
}

@article{Ortiz2017NatureHaumeaRing,
  author  = {Ortiz, J. L. and Santos-Sanz, P. and Sicardy, B. and Benedetti-Rossi, G. and B{\'e}rard, D. and Morales, N. and Duffard, R. and Braga-Ribas, F. and others},
  title   = {The size, shape, density and ring of the dwarf planet Haumea from a stellar occultation},
  journal = {Nature},
  year    = {2017},
  volume  = {550},
  pages   = {219--223},
  doi     = {10.1038/nature24051}
}

@article{Pereira2023AandA_TwoRingsQuaoar,
  author  = {Pereira, C. L. and Sicardy, B. and Morgado, B. E. and Braga-Ribas, F. and Fern{\'a}ndez-Valenzuela, E. and Souami, D. and Holler, B. J. and others},
  title   = {The two rings of (50000) Quaoar},
  journal = {Astronomy \& Astrophysics},
  year    = {2023},
  volume  = {673},
  pages   = {L4},
  doi     = {10.1051/0004-6361/202346365}
}

@article{Proudfoot2024PSJ_HaumeaJ2,
  author  = {Proudfoot, B. C. N. and Ragozzine, D. A. and Giforos, W. and Grundy, W. M. and MacDonald, M. and Oldroyd, W. J.},
  title   = {Beyond Point Masses. III. Detecting Haumea’s Nonspherical Gravitational Field},
  journal = {The Planetary Science Journal},
  year    = {2024},
  volume  = {5},
  number  = {3},
  pages   = {68},
  doi     = {10.3847/PSJ/ad26e9}
}

@article{Scott2021FrASS_AlopekeZorro,
  author  = {Scott, N. J. and Howell, S. B. and Gnilka, C. L. and Stephens, A. W. and Salinas, R. and Matson, R. A. and Furlan, E. and Horch, E. P. and Everett, M. E. and Ciardi, D. R. and Mills, D. and Quigley, E. A.},
  title   = {Twin High-Resolution, High-Speed Imagers for the Gemini Telescopes: Instrument Description and Science Verification Results},
  journal = {Frontiers in Astronomy and Space Sciences},
  year    = {2021},
  volume  = {8},
  pages   = {716560},
  doi     = {10.3389/fspas.2021.716560}
}

@article{Rabinowitz2006ApJ_HaumeaLightcurve,
  author  = {Rabinowitz, D. L. and Barkume, K. and Brown, M. E. and Roe, H. and Schwartz, M. and Tourtellotte, S. and Trujillo, C.},
  title   = {Photometric Observations Constraining the Nature of 2003 EL61},
  journal = {The Astrophysical Journal},
  year    = {2006},
  volume  = {639},
  pages   = {1238--1251},
  doi     = {10.1086/499575}
}

@article{Ortiz2023Chiron,
  author={Ortiz, J. L. and Pereira, C. L. and Sicardy, B. and Braga-Ribas, F. and others},
  title={Changing material around (2060) Chiron revealed by an occultation on December 15, 2022},
  journal={Astronomy \& Astrophysics},
  year={2023},
  volume={676},
  pages={L12},
  doi={10.1051/0004-6361/202347025}
}

@article{SantosSanz2025Char,
 author = {{Santos-Sanz}, Pablo and {Gomes-J{\'u}nior}, Altair R. and {Morgado}, Bruno E. and {Kilic}, Yucel and {Kalup}, Csilla E. and {Kiss}, Csaba and {Pereira}, Chrystian L. and {Holler}, Bryan J. and {Morales}, Nicol{\'a}s and {Ortiz}, Jos{\'e} Luis and {Sicardy}, Bruno and {Rizos}, Juan Luis and {Stansberry}, John and {French}, Richard G. and {Hammel}, Heidi B. and {Lin}, Zhong-Yi and {Souami}, Damya and {Desmars}, Josselin and {Milam}, Stefanie N. and {Braga-Ribas}, Felipe and {Assafin}, Marcelo and {Benedetti-Rossi}, Gustavo and {Camargo}, Julio I.~B. and {Duffard}, Ren{\'e} and {Rommel}, Flavia L. and {Fern{\'a}ndez-Valenzuela}, Estela and {Pinilla-Alonso}, Noem{\'\i} and {Vara-Lubiano}, M{\'o}nica},
        title = "{JWST occultation reveals unforeseen complexity in Chariklo's ring system}",
      journal = {arXiv e-prints},
         year = 2025,
        month = oct,
          eid = {arXiv:2510.06366},
        pages = {arXiv:2510.06366},
          doi = {10.48550/arXiv.2510.06366},
archivePrefix = {arXiv},
       eprint = {2510.06366},
 primaryClass = {astro-ph.EP},
       adsurl = {https://ui.adsabs.harvard.edu/abs/2025arXiv251006366S}
}

@article{Pereira2025Chiron,
  author={Pereira, C. L. and Braga-Ribas, F. and Sicardy, B. and Ortiz, J. L. and Santos-Sanz, P. and others},
  title={The Rings of (2060) Chiron: Evidence of an Evolving System},
  journal={The Astrophysical Journal Letters},
  year={2025},
  volume={992},
  pages={L19},
  doi={10.3847/2041-8213/ae0b6d}
}

@article{Rommel2023,
  author  = {Rommel, F. L. and Braga-Ribas, F. and Ortiz, J. L. and others},
  title   = {A large topographic feature on the surface of the trans-Neptunian object (307261) 2002 MS4 measured from stellar occultations},
  journal = {Astronomy \& Astrophysics},
  year    = {2023},
  volume  = {678},
  pages   = {A167},
  doi     = {10.1051/0004-6361/202346892}
}

@inproceedings{Kalup2025RingOpacity,
  author    = {Kalup, C. and Kiss, C.},
  title     = {Characterising Grains and Composition in Small-Body Ring Systems: A Case Study of Haumea's Ring},
  booktitle = {EPSC-DPS Joint Meeting 2025},
  year      = {2025},
  pages     = {EPSC-DPS2025-1768},
  doi       = {10.5194/epsc-dps2025-1768},
  note      = {Conference abstract, Helsinki, Finland, 7--12 Sep 2025}
}

@article{Kilic2020OCCPORTAL,
  author  = {Kilic, Y. and Braga-Ribas, F. and Kaplan, M. and Erece, O. and Souami, D. and Dindar, M. and Desmars, J. and Sicardy, B. and Morgado, B. E. and Shameoni, M. N. and Rommel, F. L. and Gomes-J{\'u}nior, A. R.},
  title   = {Occultation portal: A web-based platform for data collection and analysis of stellar occultations},
  journal = {Monthly Notices of the Royal Astronomical Society},
  year    = {2022},
  volume  = {515},
  number  = {1},
  pages   = {1346--1357},
  doi     = {10.1093/mnras/stac1595}
}

@article{Altmann2017HSOY,
  author  = {Altmann, M. and Roeser, S. and Demleitner, M. and Bastian, U. and Schilbach, E.},
  title   = {Hot Stuff for One Year (HSOY)},
  journal = {Astronomy \& Astrophysics},
  year    = {2017},
  volume  = {600},
  pages   = {L4},
  doi     = {10.1051/0004-6361/201730393}
}

@article{Marzari2020HaumeaRingDynamics,
  author  = {Marzari, Francesco},
  title   = {Ring dynamics around an oblate body with an inclined satellite: the case of Haumea},
  journal = {Astronomy \& Astrophysics},
  year    = {2020},
  volume  = {643},
  pages   = {A67},
  doi     = {10.1051/0004-6361/202038812}
}

@article{Roeser2010PPMXL,
  author  = {Roeser, S. and Demleitner, M. and Schilbach, E.},
  title   = {The PPMXL Catalog of Positions and Proper Motions on the ICRS. Combining USNO-B1.0 and 2MASS},
  journal = {The Astronomical Journal},
  year    = {2010},
  volume  = {139},
  number  = {6},
  pages   = {2440--2447},
  doi     = {10.1088/0004-6256/139/6/2440}
}

@misc{Arxiv2026HaumeaDeformation,
  author       = {Staelen, C. and Rambaux, N. and Chambat, F. and Castillo-Rogez, J. C.},
  title        = {Equilibrium figure of Haumea and possible detection by stellar occultation},
  year         = {2026},
  eprint       = {2603.11787},
  archivePrefix= {arXiv},
  primaryClass = {astro-ph.EP},
  doi          = {10.48550/arXiv.2603.11787},
  url          = {https://arxiv.org/abs/2603.11787}
}

\end{document}

\maketitle

\begin{figure}[p]
  \centering
  \includegraphics[width=\textwidth,height=0.88\textheight,keepaspectratio]{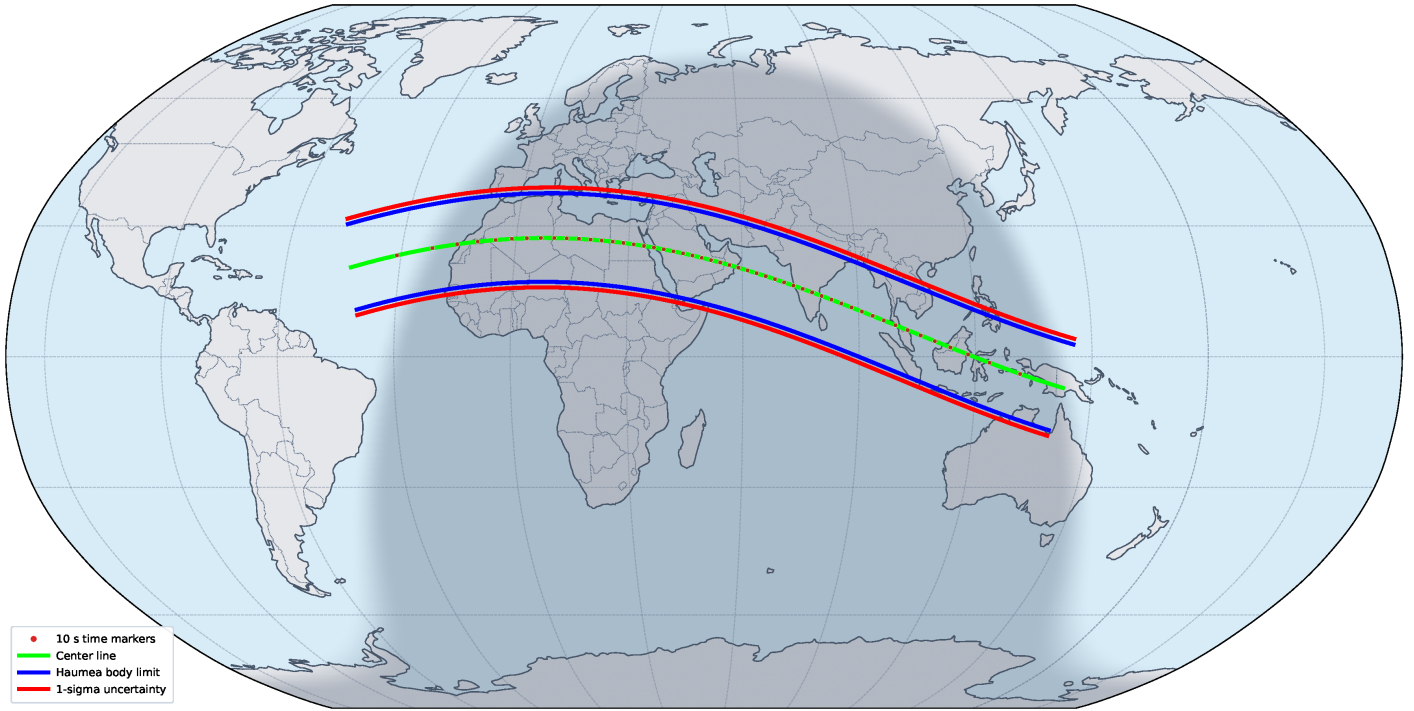}
  \caption{Stellar occultation prediction map and geometry for Haumea event 1 of 11 (Mv = 14.7). Time markers (first/middle/last): 20h 20m 40s / 20h 16m 30s / 20h 12m 30s UTC (50 markers). Note that the correct path prediction for this event is shown in Fig. 3 of the main paper.}
  \label{fig:occmap-1}
\end{figure}
\begin{figure}[p]
  \centering
  \includegraphics[width=\textwidth,height=0.88\textheight,keepaspectratio]{image5.png}
  \caption{Stellar occultation prediction map and geometry for Haumea event 2 of 11 (Mv = 19.4). Time markers (first/middle/last): 2h 11m 40s / 2h 18m 20s / 2h 24m 50s UTC (80 markers).}
  \label{fig:occmap-2}
\end{figure}
\begin{figure}[p]
  \centering
  \includegraphics[width=\textwidth,height=0.88\textheight,keepaspectratio]{image6.png}
  \caption{Stellar occultation prediction map and geometry for Haumea event 3 of 11 (Mv = 15.7). Time markers (first/middle/last): 13h 53m 0s / 13h 59m 40s / 14h 6m 20s UTC (81 markers).}
  \label{fig:occmap-3}
\end{figure}
\begin{figure}[p]
  \centering
  \includegraphics[width=\textwidth,height=0.88\textheight,keepaspectratio]{image7.png}
  \caption{Stellar occultation prediction map and geometry for Haumea event 4 of 11 (Mv = 18.3). Time markers (first/middle/last): 15h 34m 10s / 15h 38m 10s / 15h 42m 10s UTC (50 markers).}
  \label{fig:occmap-4}
\end{figure}
\begin{figure}[p]
  \centering
  \includegraphics[width=\textwidth,height=0.88\textheight,keepaspectratio]{image8.png}
  \caption{Stellar occultation prediction map and geometry for Haumea event 5 of 11 (Mv = 18.4). Time markers (first/middle/last): 12h 17m 10s / 12h 13m 30s / 12h 9m 50s UTC (45 markers).}
  \label{fig:occmap-5}
\end{figure}
\begin{figure}[p]
  \centering
  \includegraphics[width=\textwidth,height=0.88\textheight,keepaspectratio]{image9.png}
  \caption{Stellar occultation prediction map and geometry for Haumea event 6 of 11 (Mv = 19.8). Time markers (first/middle/last): 1h 8m 30s / 1h 13m 0s / 1h 17m 20s UTC (54 markers).}
  \label{fig:occmap-6}
\end{figure}
\begin{figure}[p]
  \centering
  \includegraphics[width=\textwidth,height=0.88\textheight,keepaspectratio]{image10.png}
  \caption{Stellar occultation prediction map and geometry for Haumea event 7 of 11 (Mv = 20.2). Time markers (first/middle/last): 22h 33m 20s / 22h 37m 40s / 22h 42m 0s UTC (53 markers).}
  \label{fig:occmap-7}
\end{figure}
\begin{figure}[p]
  \centering
  \includegraphics[width=\textwidth,height=0.88\textheight,keepaspectratio]{image11.png}
  \caption{Stellar occultation prediction map and geometry for Haumea event 8 of 11 (Mv = 20.4). Time markers (first/middle/last): 10h 23m 50s / 10h 27m 30s / 10h 31m 10s UTC (45 markers).}
  \label{fig:occmap-8}
\end{figure}
\begin{figure}[p]
  \centering
  \includegraphics[width=\textwidth,height=0.88\textheight,keepaspectratio]{image12.png}
  \caption{Stellar occultation prediction map and geometry for Haumea event 9 of 11 (Mv = 20.2). Time markers (first/middle/last): 14h 21m 20s / 14h 15m 50s / 14h 10m 20s UTC (67 markers).}
  \label{fig:occmap-9}
\end{figure}
\begin{figure}[p]
  \centering
  \includegraphics[width=\textwidth,height=0.88\textheight,keepaspectratio]{image13.png}
  \caption{Stellar occultation prediction map and geometry for Haumea event 10 of 11 (Mv = 20.1). Time markers (first/middle/last): 19h 10m 30s / 19h 7m 10s / 19h 3m 50s UTC (41 markers).}
  \label{fig:occmap-10}
\end{figure}
\begin{figure}[p]
  \centering
  \includegraphics[width=\textwidth,height=0.88\textheight,keepaspectratio]{image14.png}
  \caption{Stellar occultation prediction map and geometry for Haumea event 11 of 11 (Mv = 16.4). Time markers (first/middle/last): 14h 24m 30s / 14h 21m 0s / 14h 17m 40s UTC (42 markers).}
  \label{fig:occmap-11}
\end{figure}

\clearpage